\documentstyle[prb,aps]{revtex}

\begin{document}
\title{Formation and mobility of oxygen vacancies in RuSr$_{2}$GdCu$_{2}$O$_{8}$}
\author{F. Cordero$^{1}$, M. Ferretti$^{2}$, M.R. Cimberle$^{3}$, R. Masini$^{3}$}
\address{$^{1}$ CNR, Area della Ricerca di Roma - Tor Vergata, Istituto di Acustica
``O.M.Corbino``,\\
Via del Fosso del Cavaliere 100, I-00133 Roma, and INFM - Unit\`a Roma 1,
Roma, Italy}
\address{$^{2}$ LAMIA-INFM and Dipartimento di Chimica e Chimica Industriale,\\
Via Dodecaneso 31, 16146 Genova, Italy}
\address{$^{3}$ CNR-IMEM sezione di Genova c/o Dipartimento di Fisica,\\
Via Dodecaneso 33, 16146 Genova, Italy}
\maketitle

\begin{abstract}
Oxygen vacancies are introduced in the RuO$_{2}$ and possibly CuO$_{2}$
planes of RuSr$_{2}$GdCu$_{2}$O$_{8}$ by annealing in vacuum above 600~K.
The diffusive jumps of the O vacancies are accompanied by a reorientation of
the local distortion, and are probed by measuring the elastic energy loss
and modulus versus temperature at 1-10~kHz. An intense acoustic absorption
peak develops near 670~K at 1~kHz and finally stabilizes after heating up to
920~K in vacuum. The analysis of the peak shows a barrier for the O
diffusion of $\simeq 1.4$~eV, and a slowing down of Curie-Weiss type, with $%
T_{{\rm {c}}}=400-470$~K, due to the interaction among the O vacancies.

A secondary peak is attributed to O\ vacancies trapped at defects in the RuO$%
_{2}$ planes, or to vacancies in the CuO$_{2}$ planes. No sign of structural
phase transformation is found up to 920~K.
\end{abstract}

\draft


\section{INTRODUCTION}

The ruthenocuprates are a relatively new class of high-$T_{{\rm C}}$
superconductors,\cite{BWB95} which is attracting much interest for the
apparent coexistence of ferromagnetism and superconductivity.\cite{FWB99} In
spite of a considerable experimental activity on these compounds, and
notably on RuSr$_{2}$GdCu$_{2}$O$_{8}$ (Ru-1212), there are still several
obscure points regarding their preparation and characterization, and it
turns out that nominally identical samples may or may not be
superconducting, depending on subtle details of the preparation process.\cite
{LMC01c} The O stoichiometry is considered as a possible source of
non-uniform properties from sample to sample, but there is a great variety
of results regarding the influence of the preparation on the O\
stoichiometry. Some studies find that the annealing at high temperature in
vacuum or inert atmosphere causes considerable O deficiency,\cite
{BWB95,MAJ02} while others find no influence at all\cite{PTW99,CJS00} even
up to 800~$^{{\rm {o}}}$C.\cite{HFA00} It has then proposed that the
prolonged annealings affect the cation ordering,\cite{KDM00} the grain
boundaries,\cite{LMC01c} or the microstructure.\cite{TLW00} It seems that
granularity has an important role,\cite{CJS00} since the near coincidence of
the lattice parameters $a$, $b$ and $c/3$ results in the formation of small
domains with $c$ in any of the three almost equivalent directions.
Therefore, some authors attribute the improvement in superconducting
properties following long high-temperature anneals to increases in domain
size, rather than change in cation composition or O content.\cite{MZA99} The
possibility of a broad range of O stoichiometry would have implications
regarding the amount of doping of the CuO$_{2}$\ planes. In fact, from both
bond valence sum calculations\cite{MZA99} and X-ray absorption spectroscopies%
\cite{LJH01} it is found that the Ru cations in Ru-1212 exhibit mixed
valence Ru$^{4+}$/Ru$^{5+}$, with $\left[ {\rm Ru}^{4+}\right] =$ $0.4-0.5$.
The fraction of Ru$^{4+}$ is considered to be responsible for the doping of
the CuO$_{2}$ planes, with $p=\frac{1}{2}\left[ {\rm Ru}^{4+}\right] $, due
to the relative Cu/Ru stoichiometry. The value $p=0.2-0.25$ deduced in this
way, however, is much larger than that deduced from transport experiments,
suggesting that most carriers are trapped by defects or by the ferromegnetic
order.\cite{MZA99}

In the present work we address the issue whether O vacancies may be present
in Ru-1212 by anelastic spectroscopy measurements. Such a technique is
particularly suited, since, similarly to the well known case of YBa$_{2}$Cu$%
_{3}$O$_{6+x}$, an O vacancy in the RuO$_{2}$ or CuO$_{2}$ plane of Ru-1212
creates an anisotropic strain that reorients by 90$^{{\rm {o}}}$ after a
jump; therefore, the diffusive motion of the O vacancies gives rise to a
well detectable anelastic relaxation effect.

\section{EXPERIMENTAL}

Policrystalline samples of composition RuSr$_{2}$GdCu$_{2}$O$_{8}$ have been
synthesized by solid state reaction of high purity stoichiometric powders of
RuO$_{2}$, Gd$_{2}$O$_{3}$, CuO and SrCO$_{3}$, following a well established
procedure described in previous works.\cite{ACC02,ACC02b} The final step,
aging at 1070~$^{\rm{o}}$C in flowing O$_{2}$ for one week followed
cooling at 50~$^{\rm{o}}$C/h, resulted in the formation of a sintered
ingot that was cut into bars. Some samples have been characterized
structurally by X-ray powder diffraction and by magnetic susceptibility in a
Quantum Design SQUID magnetometer. The superconducting transitions presented
the two step behaviour typical of granular high-$T_{c}$ superconductors,
with intragrain onset at $T=45$~K and the intergrain one at $T=19$~K. Figure 
\ref{fig mvsT} shows the Field Cooled and Zero Field Cooled magnetization
curves measured in a field of 0.55~mT, as described in Ref. %
\onlinecite{ACC02}. The anelastic measurements were made on a bar of $%
49.5\times 4.8\times 0.58$~mm$^{3}$.

The complex Young's modulus $E\left( \omega \right) =E^{\prime }+iE^{\prime
\prime }$, was measured as a function of temperature by suspending the bar
on thin thermocouple wires and electrostatically exciting its flexural
modes. The frequencies $\omega _{i}/2\pi $ of the first three odd flexural
modes are in the ratios $1:5.4:13.3$ with $\omega _{1}/2\pi \sim 0.8$~kHz,
and are related to the real part of the Young's modulus through $\omega
_{i}=\alpha _{i}\sqrt{E^{\prime }/\rho }$, where $\alpha _{i}$ is a
geometrical factor and $\rho $ the mass density. The elastic energy loss
coefficient, or the reciprocal of the mechanical quality factor,\cite{NB72}
is $Q^{-1}\left( \omega ,T\right) =$ $E^{\prime \prime }/E^{\prime }=$ $%
S^{\prime \prime }/S^{\prime }$, where $S=S^{\prime }+iS^{\prime \prime
}=E^{-1}$ is the elastic compliance; the $Q^{-1}$ was measured from the
decay of the free oscillations or from the width of the resonance peak. An
elementary relaxation process, like the hopping of an O atom or vacancy with
rate $\tau ^{-1}$, contributes to the imaginary part of the compliance with%
\cite{NB72} 
\begin{equation}
\delta S^{\prime \prime }\left( \omega ,T\right) =\frac{cv_{0}\left( \lambda
^{(2)}-\lambda ^{(1)}\right) ^{2}}{2k_{{\rm {B}}}T}\frac{\omega \tau }{%
1+\left( \omega \tau \right) ^{2}}\,,  \label{dS"}
\end{equation}
where $\lambda ^{(i)}$ is the elastic dipole of the defect in configuration $%
i$: $c\lambda ^{(i)}$ is the anelastic strain due to a homogeneous
distribution of defects with molar concentration $c$, all in the $i$--th
configuration. Considering the relaxation from O atoms or vacancies in the
RuO$_{2}$ or CuO$_{2}$ planes, and at first neglecting the small rotation of
the RuO$_{6}$ octahedra about their $c$ axis,\cite{CJS00,MZA99} there are
only two orientations $i=1,2$ of the elastic dipole, depending whether the O
atom or vacancy (V$_{{\rm O}}$) has the nearest neighboring Ru/Cu atoms
along the $y$ or $x$ direction (corresponding to the O(1) and O(5)\ sites in
YBCO\cite{FCA90}). The upper part of Fig. \ref{fig octa} shows a projection
of the unrotated O octahedra on the $ab$ plane with the two possible types
of V$_{{\rm O}}$; $a$ and $b$ are chosen following Ref. \onlinecite{CJS00},
while the $x$ and $y$ directions are at 45$^{\rm{o}}$, parallel to the
Ru-O bonds. Strain is a centrosymmetric second rank tensor,\cite{SS82} and
can therefore be represented as an ellipsoid with the principal axes
representing the deviations of the lengths with respect to the reference
unstrained sphere (dotted circles).\cite{note} The ellipsoids in Fig. \ref
{fig octa} represent contractions around the V$_{{\rm O}}$, more pronounced
along the directions of the nearest neighbor Ru atoms. The lower part of the
figure includes the (slightly exaggerated) rotations of the octahedra; the
symmetry of the elastic dipoles, however, remains unaltered, as can be seen
from the mirror planes passing through the V$_{{\rm O}}$, which are still
perpendicular to the $x$ or $y$ directions. Analogously to the case of YBCO,%
\cite{BSW94} if we call $\lambda _{1}=\lambda _{xx}^{(1)}$ and $\lambda
_{2}=\lambda _{yy}^{(1)}$, then it is $\lambda _{xx}^{(2)}=\lambda _{2}$ and 
$\lambda _{yy}^{(2)}=\lambda _{1}$, while $\lambda _{zz}^{(1)}=\lambda
_{zz}^{(2)}$ and $\lambda _{\mu \nu }^{(i)}=0$ for $\mu \neq \nu $. The
anelastic strain associated with an O jump is therefore $\varepsilon
_{1}-\varepsilon _{2}$, and causes relaxation of the $\delta \left(
S_{11}-S_{12}\right) $ compliance, with $\lambda _{1}-\lambda _{2}$ entering
into eq. (\ref{dS"}); the factor $1/2$ in that expression is appropriate for
the case of a $\left\langle 100\right\rangle $ orthorhombic elastic dipole
in a tetragonal crystal.\cite{NB72} The above expression is peaked at $%
\omega \tau =1$, and since the measurements are made at the resonance
frequencies $\omega _{i}$ as a function of temperature, one finds peaks at
the temperatures $T_{i}$ such that $\omega _{i}\tau \left( T_{i}\right) =1$.
Therefore, measurements at different frequencies provide the temperature
dependence of the rate $\tau ^{-1}\left( T\right) $, which generally follows
the Arrhenius law: 
\begin{equation}
\tau ^{-1}\left( T\right) =\tau _{0}^{-1}\exp \left( -W/k_{{\rm {B}}%
}T\right) \,,  \label{Arrh}
\end{equation}
where $W$\ is the energy barrier for the O jump. Assuming that $E^{\prime }$
and $S^{\prime }$ are almost constant over the temperature range in which
the peak is observed, eq. (\ref{dS"}) describes also the peaks in $E^{\prime
\prime }\left( \omega ,T\right) $ and $Q^{-1}\left( \omega ,T\right) $: 
\begin{equation}
Q^{-1}\simeq \Delta \frac{\omega \tau }{1+\left( \omega \tau \right) ^{2}}\,,
\label{Q-1}
\end{equation}
where the relaxation strength 
\begin{equation}
\Delta \left( T\right) =\frac{\delta S^{\prime \prime }\left( 0,T\right) }{%
S^{\prime \prime }}=\frac{cv_{0}\left( \lambda ^{(2)}-\lambda ^{(1)}\right)
^{2}}{S^{\prime }\,k_{{\rm {B}}}T}  \label{Delta}
\end{equation}
corresponds to the static elastic susceptibility. A process characterized by
a spectrum of relaxation times may be described by the Fuoss-Kirkwood
distribution of relaxation rates,\cite{NB72} yielding for the frequency
dispersion factor $\left[ 1+\left( i\omega \tau \right) ^{\alpha }\right]
^{-1}$ instead of $\left[ 1+i\omega \tau \right] ^{-1}$; the imaginary part
of the dispersion to be inserted into eqs. (\ref{dS"}) and (\ref{Q-1})
therefore becomes 
\begin{equation}
\frac{\omega \tau }{1+\left( \omega \tau \right) ^{2}}\rightarrow \frac{%
\left( \omega \tau \right) ^{\alpha }\sin \left( \frac{\pi }{2}\alpha
\right) }{1+\left( \omega \tau \right) ^{2\alpha }+2\left( \omega \tau
\right) ^{\alpha }\cos \left( \frac{\pi }{2}\alpha \right) }\,.  \label{FK}
\end{equation}

\section{RESULTS}

Figure \ref{fig spectra} presents three measurement runs in vacuum up to
660, 850 and 920~K; the lower panel contains the elastic energy loss
coefficient, while the upper panel contains the relative change of the
Young's modulus, where $E_{0}=E^{\prime }\left( T=0\right) $ refers to the
as prepared state. During the first heating the dissipation is low, and
starts increasing above 600~K (curve 1). This increase is due to the
formation of O vacancies, as confirmed by heating the same sample in a UHV
system equipped with a residual gas analyzer; the trace of the O$_{2}$
partial pressure started increasing at 600~K, indicating that this is the
temperature of the onset of massive O loss. During the subsequent heating
(curve 2 at $2-3$~K/min) a $Q^{-1}\left( T\right) $ peak develops and
becomes stable after heating above $\sim 850$~K. In fact, the $Q^{-1}\left(
T\right) $ curve measured on cooling (curve 3, open circles) is retraced
during the subsequent runs: curve 4 on heating is slightly higher, but curve
5 on cooling (crosses) perfectly coincides with curve 3. This means that no
further loss of O occurs, at least up to 920~K in vacuum. This fact again is
confirmed by outgassing the same sample in the UHV system, and noting that
the O outgassing rate at 900~K drops by two orders of magnitude within
10~min. The amount of O loss has been estimated from the mass variation
after oxygenation and outgassing treatments, and ranges between 2.2 and 3\%
per formula unit after outgassing at 900-1000~K for less than 1~h.

From the perfect coincidence of the $Q^{-1}\left( T\right) $ curves 3 and 5
measured on cooling, also very close to curve 4 on heating, we conclude that
curves 3--5 represent a situation very close to equilibrium and their
analysis is meaningful. The $E^{\prime }\left( T\right) $ curves reflect the
behavior of the acoustic losses, with a negative step in correspondence to
the absorption peak.\cite{NB72} A transformation to a more symmetric
structure where the rotation angle of the RuO$_{6}$ octahedra about their $c$
axes goes to zero may be expected at high temperature,\cite{CJS00} but there
is no trace of it from room temperature to 930~K. Such a transformation
should appear in the $E^{\prime }\left( T\right) $ curves as a jump, dip or
change of slope.

\section{DISCUSSION}

The formation of O vacancies in the RuO$_{2}$ and possibly CuO$_{2}$ planes
in Ru-1212 should have analogies with the partial filling of the CuO$_{x}$
planes in YBa$_{2}$Cu$_{3}$O$_{6+x}$. From the structural point of view the
CuO$_{2}$ planes of Ru-1212 correspond to the CuO$_{2}$ planes of YBCO,
which remain stoichiometric, while the RuO$_{2}$ planes of Ru-1212
correspond to the CuO$_{x}$ planes of YBCO.\cite{CJS00} Based on this
structural analogy, we assume that the O\ vacancies form mainly in the RuO$%
_{2}$ planes. It is also plausible that such planes may accommodate V$_{{\rm %
O}}$, in view of the mixed valence displayed by Ru both in Ru-1212\cite
{LJH01} and Ru-1222;\cite{WJL02,AKY02b} in the latter compound, the O
stoichiometry seems to affect the Ru valence,\cite{AKY02b} again supporting
the hypothesis of V$_{{\rm O}}$ in the Ru planes.

\subsection{Oxygen hopping in YBCO}

The case of YBCO is complicated, since only half filling of the available O
sites can be achieved, with the formation of parallel Cu-O\ chains in the
ortho-I phase. A\ complex $x-T$ phase diagram results, which can be
reproduced, for example, introducing three different short range O-O
interaction energies (so called ASYNNNI model).\cite{FCA90,MLA01} The
diffusive jumps of the O atoms in the CuO$_{x}$ planes in the ortho-I phase
have been studied by anelastic relaxation,\cite{XCW89,CS91,SHW92,BSW94}
finding that hopping occurs over a barrier of $\sim 1.1$~eV; it has also
been shown that there are actually two additional distinct hopping rates,
with smaller activation energies, for O atoms belonging to chains in the
ortho-II and tetragonal phases\cite{33,23} and for isolated O atoms.\cite
{33,46} In spite of the numerous investigations, the anelastic spectrum due
to O hopping in YBCO has not been quantitatively explained in terms of the
microscopic parameters, like the elastic dipole associated to an O atom and
the short-range O-O interaction energies, e.g. of the ASYNNNI model. Still,
it has been shown that the relaxation strength (the static elastic
susceptibility)\ exhibits a Curie-Weiss enhancement and the relaxation rate
a critical slowing near the orthorhombic/tetragonal phase transformation,%
\cite{BSW94} corresponding to the ordering of O in the O(1) sublattice. The
anisotropy of the elastic dipole in YBCO is found to be $\left| \lambda
_{1}-\lambda _{2}\right| \simeq 0.02$ from the dependence of the cell
parameters $a$ and $b$ on the O content both in the orthorhombic and
tetragonal phases.\cite{BFW97} The same information can hardly be obtained
from anelastic measurements on ceramic samples, due to the heavy corrections
necessary for the angular averaging of the elastic constants and the
porosity of the samples.\cite{SLN95}

\subsection{O vacancies and related defects in Ru-1212}

The case of Ru-1212 should be simpler, since the RuO$_{2}$ planes remain
close to the full stoichiometry even after prolonged outgassing; this can be
described in terms of a deficiency $\delta $ in RuO$_{2-\delta }$ with $%
\delta <0.03$, as deduced from the mass variation after
oxygenation/outgassing treatments. This means that one can consider the
dynamics of isolated or at most paired O vacancies with small concentration $%
c=\delta $, instead of more complex chain structures. One can therefore
treat the O vacancy exactly like an O atom in a nearly empty CuO$_{x}$
plane, and with an elastic dipole $\lambda ^{{\rm {V}}}=-\lambda ^{{\rm {O}}%
} $.

A complication arises from the fact that the main peak at $\sim 670$~K
(1~kHz) is accompanied by a shoulder at $\sim 530$~K, indicating that there
are actually two distinct relaxation processes associated with V$_{{\rm O}}$
in Ru-1212. There are various explanations for two distinct relaxation
processes associated with V$_{{\rm O}}$. One is that V$_{{\rm O}}$ may form
stable pairs with a binding energy $-E_{b}$; in that case, the reorientation
of the vacancy pair would require its temporary dissociation and the
corresponding relaxation rate should have an activation energy increased by $%
E_{b}$ over the barrier for hopping of a simple V$_{{\rm O}}$. The
corresponding anelastic relaxation peak would appear at higher temperature,
analogously to the case of the diluted interstitial O atoms and O pairs in La%
$_{2}$CuO$_{4+\delta }$.\cite{63} The problem with this interpretation is
that the peak at lower temperature in Fig. \ref{fig spectra}, which should
be due to isolated V$_{{\rm O}}$, has definitely a smaller intensity than
the peak at higher temperature. At these low concentrations, $\delta <0.03$,
one would expect on the contrary that the isolated V$_{{\rm O}}$ predominate
over the V$_{{\rm O}}$ pairs, rendering this explanation unlikely.

Another possibility is that the peak at lower temperature is associated with
the complex of a V$_{{\rm O}}$ with a defect, for example a Cu atom
substituting a Ru atom (Cu$_{{\rm Ru}}$). The fact that the temperature of
the Cu$_{{\rm Ru}}-$V$_{{\rm O}}$ relaxation is lower than that of V$_{{\rm O%
}}$ may be understood in terms of a Cu$_{{\rm Ru}}$ environment providing a
lower barrier for O hopping, in accordance with the observation that the
barrier for O\ hopping in the CuO$_{x}$ planes of YBCO is 1.1~eV, lower than
the barrier associated with the main peak in Ru-1212 (see later on). Support
to the hypothesis of the existence of Cu$_{{\rm Ru}}$ defects comes from the
fact that Ru$_{1-x}$Cu$_{x}$Sr$_{2}$GdCu$_{2}$O$_{8}$ with mixed Ru$_{1-x}$Cu%
$_{x}$O$_{2}$ planes can actually be synthesized and is superconducting.\cite
{KDK01} In addition, superlattice spots in electron diffraction patterns of
Ru-1212 have been interpreted in terms of domains with intermixed and
ordered Cu/Ru ions in the nominal RuO$_{2}$ planes,\cite{AIN02} or,
alternatively, with ordering of mixed valence Ru$^{4+}$/Ru$^{5+}$ ions.\cite
{AIN02} The presence of the minor peak in the anelastic spectra may be
accounted for by Cu$_{{\rm Ru}}$ defects dispersed at a concentration much
smaller than that of the V$_{{\rm O}}$, {\it i.e. }$<1\%$, and acting as
trapping centers for few V$_{{\rm O}}$, but also by the minority regions of
mixed Cu/Ru or also Ru$^{4+}$/Ru$^{5+}$ devised in Ref. \onlinecite{AIN02}.
Note that Cu$_{{\rm Ru}}$ defects in the RuO$_{2}$ planes require an equal
amount Ru$_{{\rm Cu}}$ in the CuO$_{2}$ planes, or the presence of impurity
phases rich in Ru.

It is also possible to explain the minor peak in the anelastic spectra
without invoking Cu/Ru intermixing, but supposing that that V$_{{\rm O}}$
can form also in the CuO$_{2}$ planes of Ru-1212, at concentrations lower
than in the RuO$_{2}$ planes. In this case the peak at lower temperature
would be due to the minority of faster V$_{{\rm O}}$ in the CuO$_{2}$
planes. In what follows we will focus on the main peak.

\subsection{Fit of the relaxation curves and Curie-Weiss response}

Attempts to fit the main peak in terms of Eqs. (\ref{dS"}--\ref{FK}) yield a
mean relaxation rate with $\tau _{0}^{-1}\simeq 4\times 10^{17}$~s$^{-1}$
and $E=1.9$~eV, with a broadening parameter $\alpha =0.7$; such a value for $%
\tau _{0}^{-1}$ is too high for an O vacancy or any defect. In addition, it
is impossible to reproduce the marked decrease of the peak intensity with
increasing temperature (see the curve at higher frequency in Fig. \ref{fig
fit}). These features suggests a Curie-Weiss like enhancement of the
relaxation strength, like in YBCO,\cite{BSW94} due to the long range elastic
interactions among the V$_{{\rm O}}$. The effect of the long range elastic
interaction among elastic dipoles has been treated in connection with the O\
atoms in YBCO\cite{BSW94} and also the interstitial O atoms in bcc metals.%
\cite{DBR82,HWD92} The interaction causes a tendency to cooperative
alignment of the dipoles and results in a factor $\left( 1-T_{{\rm C}%
}/T\right) ^{-1}$ that multiplies both the relaxation strength $\Delta $ and
time $\tau $, where $T_{{\rm C}}\propto c\left( 1-c\right) $ is the
temperature below which the elastic dipoles at concentration $c$ start
aligning in a same direction. Note that the elastic dipoles can describe
equivalently O atoms or vacancies, with opposite sign, and in the present
situation it is natural to refer to vacancies. The relaxation process is
then decribed by 
\begin{equation}
dS^{\prime \prime }=S^{\prime }\,\,\tilde{\Delta}\frac{\omega \tilde{\tau}}{%
1+\left( \omega \tilde{\tau}\right) ^{2}}\,  \label{Q-1 CW}
\end{equation}
\[
\tilde{\Delta}=\frac{cv_{0}\left( \lambda _{1}-\lambda _{2}\right) ^{2}}{%
S^{\prime }\,k_{{\rm {B}}}\left( T-T_{{\rm C}}\right) }\,\,,\quad \tau
^{\prime }=\frac{\tau }{1-T_{{\rm C}}/T}\,. 
\]
The divergence of the relaxation time on approaching $T_{{\rm C}}$ from
above results in a $\tau ^{-1}\left( T\right) $ that is steeper than Eq. (%
\ref{Arrh}), and therefore in a higher apparent energy barrier $W$ and
prefactor $\tau _{0}^{-1}$, as presently observed. The divergence in the
relaxation strength results in a peak intensity increasing more than as $1/T$
with decreasing $T$.

Figure \ref{fig fit} presents a fit with Eq. (\ref{Q-1 CW}), without relying
on the usual approximation that the elastic compliance $S^{\prime }$ or
modulus $E^{\prime }$ and therefore also the resonance frequencies $\omega
_{i}\propto \sqrt{E^{\prime }}$ are constant. Therefore, instead of $Q^{-1}$%
, Fig. \ref{fig fit} reports $S^{\prime \prime }\left( \omega ,T\right)
/S_{0}=$ $Q^{-1}\left( \omega ,T\right) \,S^{\prime }\left( \omega ,T\right)
/S_{0}$ with $S_{0}=S^{\prime }\left( \omega ,0\right) $ and $S^{\prime
}\left( \omega ,T\right) /S_{0}=\left[ \omega _{i}\left( T\right) /\omega
_{i}\left( 0\right) \right] ^{-1/2}$. This correction for the temperature
dependence of the frequency and elastic compliance improves slightly the
fit. The dashed curve is obtained with $W=1.33$~eV, $\tau _{0}=1\times
10^{-14}$~s, $T_{{\rm C}}=460$~K for the main peak and $W=1.4$~eV, $\tau
_{0}=2\times 10^{-17}$~s, $\alpha =0.63$ for the smaller peak. The fit is
not perfect, but the main process can be described with very reasonable
parameters; in addition, there is no need of introducing an additional
broadening, since the theoretical curve has already the correct width. The
introduction of broadening, e.g. through the Fuoss-Kirkwood distribution
with Eq. (\ref{FK}), introduces the additional parameter $\alpha $ and
allows slightly better fits to be obtained, like the continuous curve with $%
W=1.46$~eV, $\tau _{0}=1.1\times 10^{-15}$~s, $T_{{\rm C}}=470$~K, $\alpha
=0.82$ for the main peak and $W=1.10$~eV, $\tau _{0}=2\times 10^{-14}$~s, $%
\alpha =0.7$ for the smaller peak. A\ probable origin of the peak broadening
is the high density of antiphase boundaries between domains with opposite
senses of rotation of the octahedra.\cite{MZA99,CJS00} It appears, however,
that the introduction of a broadening is not the main ingredient for
improving the accordance of the data with the Curie-Weiss expression, Eq. (%
\ref{Q-1 CW}); possibly the mean field treatment\cite{DBR82} of the
interactions among the elastic dipoles is not adequate, or some changes of
the electronic or structural properties of Ru-1212 make the barrier for O
hopping temperature dependent, over the wide range of temperature
(500-900~K) where the peak is observed. For example, the angle of rotation
of the RuO$_{6}$ octahedra about their $c$ axes has been found to decrease
almost linearly when temperature increases from 0 to 300~K,\cite{CJS00} and
it might well continue decreasing at higher temperatures. Still, from the
fits shown in Fig. \ref{fig fit} and others not reported here, it appears
that the main peak is due to hopping of the O vacancies with an activation
energy of 1.3--1.4~eV; in addition, there is considerable critical slowing
of the hopping dynamics and enhancement of the elastic susceptibility on
approaching an ordering temperature $T_{{\rm C}}=400-470$~K. The ordering
would consist in the preferential occupation of sites of one type, namely
with the nearest neighbor Ru atoms along either $x$ or $y$ (see Fig. \ref
{fig octa}), and would result in a slightly orthorhombic cell. It would be
similar to the tetragonal-to-orthorhombic transformation in YBCO, where O\
occupies preferentially the O(1) instead of the O(5) sites, but not
necessarily with the formation of chains of vacancies.

The minor relaxation at lower temperature is almost completely masked by the
main relaxation process, and the fit does not yield reliable values of the
parameters.

\subsection{Relaxation strength and comparison with YBCO}

A reliable estimate of the shape factor $\left| \lambda _{1}-\lambda
_{2}\right| $ of the elastic dipole tensor associated with an O vacancy
cannot be done at present. In fact, Eq. (\ref{dS"}) refers to a single
crystal excited on pure $\varepsilon _{1}-\varepsilon _{2}$ mode, while a
ceramic sample requires averaging of the elastic constants over the solid
angle and corrections due to porosity.\cite{SLN95} The same is true for
YBCO, were the shape factor can be estimated from the structural data at
different O contents ($\left| \lambda _{1}-\lambda _{2}\right| =$ 0.19 in
Ref. \onlinecite{BFW97} or 0.25 from the data in Ref. \onlinecite{CSH96}). A
comparison between Ru-1212 and YBCO in the ortho-I phase\cite
{BSW94,XCW89,CS91,SHW92,33} shows that the peaks due to O hopping have
comparable intensities in both materials. This may at first indicate that
Ru-1212 has much larger $\left| \lambda _{1}-\lambda _{2}\right| $ than
YBCO, since in the present case the concentration of elastic dipoles is $%
<0.03$ (the concentration of vacancies), while in YBa$_{2}$Cu$_{3}$O$_{\sim
6.9}$ the molar concentration of mobile O atoms or vacancies is $\sim 1$. It
should be considered, however, that in the case of orthorhombic YBCO, the O
atoms jump between sites that are energetically inequivalent, and this
reduces the relaxation strength by a factor\cite{35} $4n_{x}n_{y}$, where $%
n_{x}$ and $n_{y}$ are the equilibrium fractions of occupied sites of type $%
x $ or $y$; in the ortho-I phase it is $n_{x}\ll n_{y}$ and therefore $%
n_{x}n_{y}\ll 1$. The relatively small amplitude of the peak in the ortho-I
phase of YBCO should therefore be attributed to this effect; indeed, the
relaxation strength reaches values as high as 1 near the
tetragonal-to-orthorhombic transition of YBCO,\cite{BSW94} where $%
n_{x}\simeq n_{y}\simeq 1/2$.

An estimate of the order of magnitude of $\left| \lambda _{1}-\lambda
_{2}\right| $ in Ru-1212 can be made as follows: the polycrystalline average
of the reciprocal Young's modulus, $E^{-1}$, assuming uniform stress from
grain to grain of a tetragonal crystal (a different approximation would be
that of uniform strain), is\cite{NB72,SS82} 
\begin{equation}
\left\langle E^{-1}\right\rangle =\frac{1}{5}\left( 2S_{11}+S_{33}\right) +%
\frac{1}{9}\left( 2S_{44}+S_{66}+4S_{13}+2S_{12}\right) \,,
\end{equation}
of which only $\delta \left( S_{11}-S_{12}\right) $ relaxes according to Eq.~%
\ref{dS"}; therefore\cite{NH65} $\delta \left\langle E^{-1}\right\rangle =%
\frac{4}{45}\delta \left( S_{11}-S_{12}\right) $ and, assuming that the
porosity affects in the same way $\delta \left\langle E^{-1}\right\rangle $
and $\left\langle E^{-1}\right\rangle $, the relaxation strength becomes
(including the Curie-Weiss correction) 
\begin{equation}
\Delta =\frac{\delta E^{-1}}{E^{-1}}=\frac{2}{45E^{-1}}\frac{cv_{0}}{k_{{\rm 
{B}}}\left( T-T_{{\rm C}}\right) }\left( \lambda _{1}-\lambda _{2}\right)
^{2}\,.  \label{Delta2}
\end{equation}
The effective Young's modulus of our sample was $E\sim 40$~GPa (the low
value is due to a porosity of nearly 50\%), $v_{0}=170$~\AA $^{3}$ and
assuming $c=0.02$, and $\Delta =0.022$ at $680~$K from the fit of Fig. \ref
{fig fit} with $T_{{\rm C}}=460$~K, we find $\left| \lambda _{1}-\lambda
_{2}\right| \simeq 0.10$. There are several sources of error in such an
estimate, from the polycrystalline average to the V$_{{\rm O}}$
concentration, but 0.1 is a very reasonable value for the shape factor of
the V$_{{\rm O}}$; it is $4-5$ times larger than the value estimated for
YBCO, but $5-10$ times smaller than the shape factor of interstitial O, N
and C atoms in bcc metals.\cite{NB72} It can be concluded that the
high-temperature relaxation processes in Ru-1212 can consistently be
interpreted in terms of hopping of V$_{{\rm O}}$, with a shape factor
somewhat larger than in YBCO.

\subsection{Possible influence of the O vacancies in the transport properties of Ru-1212}

The flat high-temperature anelastic spectrum of Ru-1212, after
the initial oxygenation treatment, indicates absence of O\ vacancies or
interstitial O atoms; it can be safely stated that the concentration of such
O defects is at least 10 times smaller than after O outgassing, namely of
the order of 0.2\% or less. This means that the level of the O\ vacancies
may be kept as low as $\delta \le 0.002$ also without treatments at high O$%
_{2}$ pressures, and should not exceed in any case $\delta =0.02-0.03$ . The
latter values are estimated from the mass variation after
oxygenation/outgassing treatments up to 1000~K, and correspond to a rather
stable concentration of V$_{{\rm O}}$, as deduced from the stability of the
anelastic spectra during aging in vacuum at high temperature. Unless
assuming that a V$_{{\rm O}}$ may neutralize or trap more than two holes,
even a concentration $\delta \sim 0.03$ of V$_{{\rm O}}$ would not explain
the discrepancy between the hole concentration deduced from the valence of
Ru and that deduced from transport experiments.\cite{MZA99} The present
study, therefore, even though demonstrates the existence of O\ vacancies in
Ru-1212, supports the view that the origin for the poor superconducting
properties often found in Ru-1212 is not in the O stoichiometry.

\section{CONCLUSION}

It has been shown that RuSr$_{2}$GdCu$_{2}$O$_{8-\delta }$ may be prepared
with very low ($\delta <0.2\%$) O deficiency, also without treatments in
high O$_{2}$ pressure, and annealing in vacuum above 600~K introduces O
vacancies in the RuO$_{2}$ and/or CuO$_{2}$ planes. The structural analogy
with the cuprates of the YBCO family suggests that most of the vacancies are
formed in the RuO$_{2}$ planes. The diffusive motion of these vacancies
gives rise to an intense anelastic relaxation process, measured as a peak of
the elastic energy loss coefficient around 670~K at 1~kHz. The concentration
of the O vacancies reaches a plateau estimated as $\delta =0.02-0.03$ from
the mass variation after treatments up to 1000~K in vacuum or O$_{2}$
atmosphere. The analysis of the peak shows that the barrier for the O
diffusion is of $\simeq 1.4$~eV, and the long range elastic interaction
among the O\ vacancies causes an enhancement of the relaxation strength and
time of Curie-Weiss type, with $T_{{\rm C}}=400-470$~K.

A minor peak at $\sim 530$~K might be due to O\ vacancies trapped at
defects, like Cu substituting Ru in the RuO$_{2}$ planes, or to vacancies in
the CuO$_{2}$ planes. No sign of structural phase transformation appears in
the elastic modulus and absorption curves versus temperature up to 930~K.


\begin{references}
\bibitem{note}  The elastic dipole is actually a quadrupole, and is called
''dipole'' due to its role in anelastic relaxation, which is parallel to the
role of the electric dipole in dielectric relaxation.\cite{NH65}
\end{references}

\section{Figures}

\begin{figure}[tbp]
\caption{Field Cooled (closed circles) and Zero Field Cooled (open circles)
magnetization of an as-prepared sample at an external field of 0.55~mT.}
\label{fig mvsT}
\end{figure}

\begin{figure}[tbp]
\caption{Upper part: view of the RuO$_2$ plane with the RuO$_6$ octahedra
put in evidence (the apical O atoms are not shown). The ellipsoids represent
the elastic dipoles associated with the O vacancies. Lower part: the same
view with the octahedra rotated about their $c$ axes.}
\label{fig octa}
\end{figure}

\begin{figure}[tbp]
\caption{Anelastic spectrum of Ru-1212 measured at 0.8~kHz during three
heating and cooling cycles, in the sequence indicated by the numbers. Lower
panel: elastic energy loss coefficient; upper panel: relative variation of
the Young's modulus.}
\label{fig spectra}
\end{figure}

\begin{figure}[tbp]
\caption{Elastic energy loss of Ru-1212 measured during the last cooling
cycle (curve 5 of Fig. 1), measured at two frequencies. The lines are fits
as described in the text.}
\label{fig fit}
\end{figure}

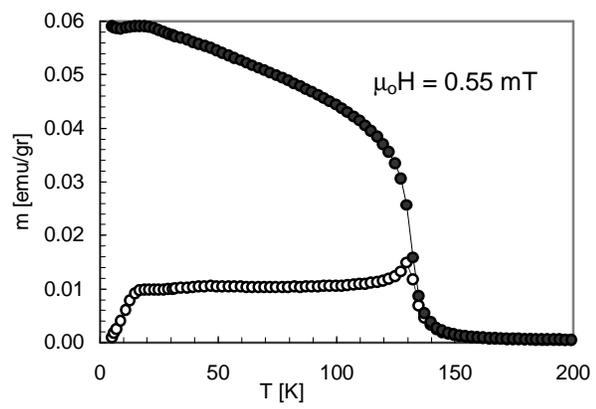
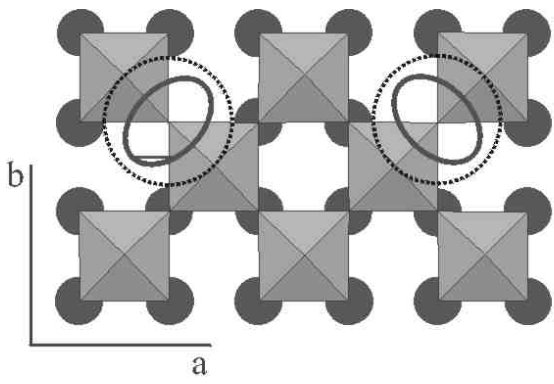
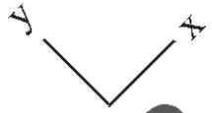
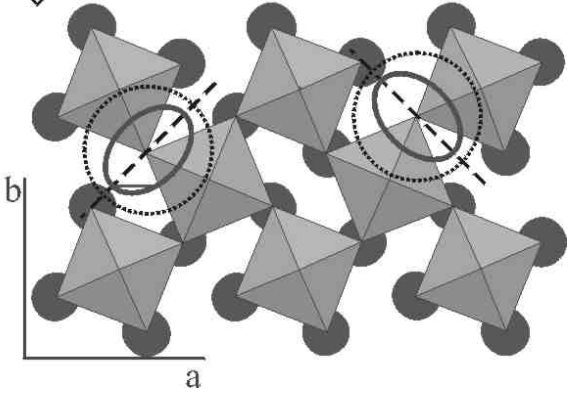
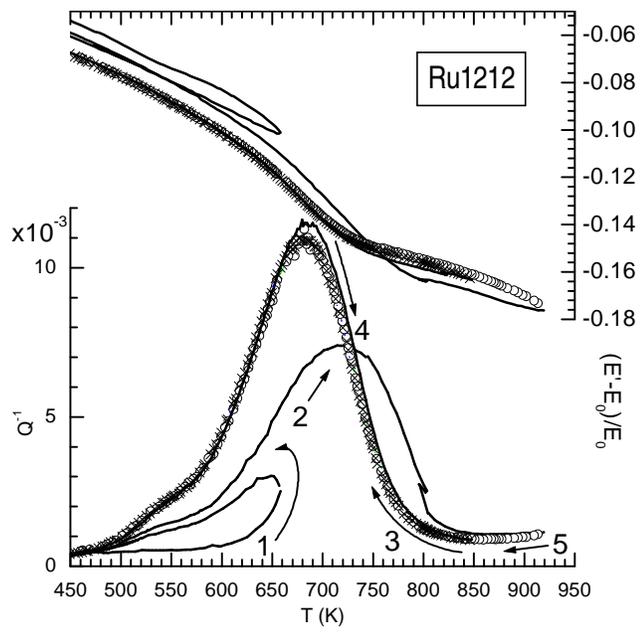
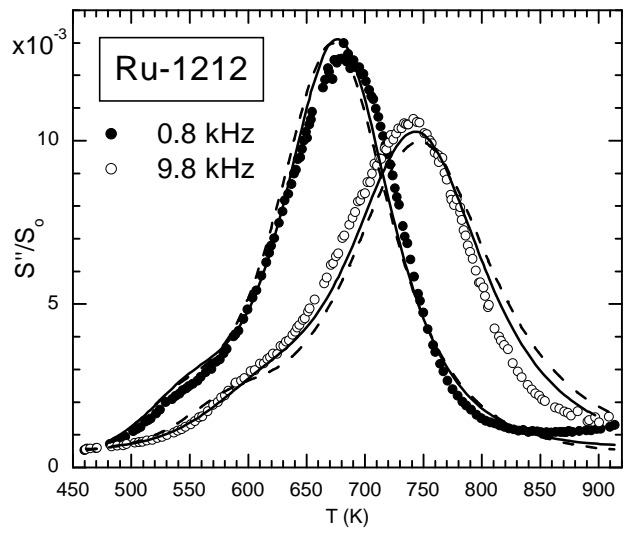
\end{document}